\documentclass[aps,twocolumn,prb]{revtex4}
\usepackage{graphicx}
\usepackage{epsfig}
\usepackage{amsmath}
\usepackage{graphics}

\begin{document}

\title{Self-similarity of single-channel transmission 
for electron transport in nanowires}

\author{M.F. Gelin, Zhenyu Li and D.S. Kosov }

\address{Department of Chemistry and Biochemistry, University of Maryland,
College Park, MD 20742}

\begin{abstract}
We demonstrate that the single-channel transmission in the resonance
tunneling regime exhibits self-similarity as a function of the nanowire
length and the energy of incident electrons. 
The self-similarity is used to design the nonlinear
transformation of the nanowire length and 
energy which, on the basis
of known values of transmission for a certain region on the energy-length
plane, yields transmissions for other regions on this plane.
Test calculations with a one-dimensional tight-binding model illustrate the described
transformations.
Density function theory based transport calculations of Na atomic wires 
 confirm the existence of the self-similarity in the transmission.
\end{abstract}
\maketitle

\section{Introduction}

When a nanowire is attached to two electron reservoirs with different
chemical potentials, the electric current flows through it. Studies
of the electrical conductance at the nanoscale have resulted in
discoveries of interesting physical effects, which include conductance
quantization, molecular rectification, negative differential resistance,
hysteresis, oscillatory length-dependence of the conductance.\cite{agrait03,datta04}
These and other unusual and potentially useful properties make nanowires
promising candidates for creating various electronic nanodevices.\cite{nitzan03}

Electrical conduction in nanowires can be approximated under certain conditions as
a ballistic scattering of electrons along the one-dimensional quasiperiodic superlattice 
of finite length. Since the first fabrication
of quasi-periodic $GaAs-AlAs$ multilayer structure,\cite{Mer_85}
propagation of elementary excitations (electrons, photons, polaritons,
spin waves) through one-dimensional fractal structures has been extensively 
studied.\cite{Cott_03} It has been demonstrated, in particularly, that the
quasi-periodic structure of the medium manifests itself in the self-similarity
of the transmission.\cite{Gap_02,Gap_04,Alb_99,Gong_98,Mac_97} The
Landauer conductance of the generalized Thue-Morse and Fibonacci one-dimensional
lattices has also been investigated.\cite{Roy_95a,Roy_95b,Sal_02,Sal_98,Wang_00,Wang_01,Claro_02}
In addition, the fractal conductance fluctuations have been predicted
and measured in two-dimensional electronic billiards in magnetic fields.\cite{2D_bil}
If the system has a
quasiperiodic structure and consists of a number of identical or similar scattering blocks, 
the transmission through a single block can be mapped into the transmission
through the entire system. This is the central idea behind several powerful approaches to the 
calculation of transmission through quasiperodic structures, those like  
transfer matrix \cite{per02} and Dyson-equation.\cite{Oni_99}  The aim of our paper is
to explore the possibility of the self-similarity for single-channel transmission 
for electron transport in nanowires.

We demonstrate that a single-channel transmission
in the resonance scattering regime exhibits self-similarity as a function
of the incident electron 
energy and the nanowire length. We use one-dimensional tight-binding
model with nearest-neighbor hopping and Landauer theory to explain
the self-similarity of the transmission coefficient for electron transport
in nanowires. Based upon resonance condition for
the  transmission, we derive  analytical formulas for nonlinear
continuous transformation of the wire length and the energy of the transmitted electrons. 
The transformation preserves
the number of transmission maxima by deforming initial region on the
energy -- wire length plane. We confirm the existence of the self-similarity
by numerical calculation of the transmission for various wire lengths,  
as well as by comparison with the first principle transport
 calculations
for sodium wires.

\section{Self-similarity of transmission }

To describe a nanowire of length $N$, we adopt the standard tight-binding
model with the nearest neighbor constant hopping rate $V$ and the
site energy $E_{0}$. The system Hamiltonian is thus described by
the three-diagonal $N\times N$ matrix:
\begin{equation}
\mathbf{H}=\left[\begin{array}{ccccccc}
E_{0} & V & 0 & 0 & 0 & 0 & 0\\
V & E_{0} & V & 0 & 0 & 0 & 0\\
0 & V & E_{0} & V & 0 & 0 & 0\\
0 & 0 & . & . & . & 0 & 0\\
0 & 0 & 0 & . & . & . & 0\\
0 & 0 & 0 & 0 & V & E_{0} & V\\
0 & 0 & 0 & 0 & 0 & V & E_{0}
\end{array}\right].
\label{H}
\end{equation}
 The interaction between the wire and the leads is not considered
explicitly and is accounted for by the introduction of the self-energy
matrix, $\mathbf{\Sigma}(E)$, into the wire Green's function, \begin{equation}
\mathbf{G}(E)=[E-\mathbf{H}-\mathbf{\Sigma}(E)]^{-1}.\label{Gr}\end{equation}
 Assuming that the molecule is connected to the left (right) lead through
its $1(N)$th sites, the transmission is determined by the $1N$
element of the Green's function (\ref{Gr}):\cite{Muj94a,Muj94b}

\begin{equation}
T(E)=4\Delta^{2}(E)|G_{1N}(E)|^{2},\label{T}\end{equation}
 where $\Delta(E)$ is the imaginary part of the self-energy function
\begin{equation}
\Sigma(E)=\Lambda(E)-i\Delta(E).
\label{self}
\end{equation}
 The conductance at low temperature and small voltage is given (in
units of $e^{2}/\pi\hbar$) by the  Landauer formula
\begin{equation}
g=T(E_{F}),
\label{g}
\end{equation}
 $E_{F}$ being the Fermi-energy of the electrodes.

The matrix element of the molecular Green's function $G_{1N}(E)$ 
can be analytically computed for our model
Hamiltonian.\cite{Muj94b,Muj94a} To simplify the presentation,
we define the self-energy within the broadband approximation assuming
that $\Lambda=0$ in eq.(\ref{self}) and considering $\Delta$ as
an energy-independent constant.\cite{Muj94b} The assumption can be
relaxed and its consequences are discussed later. 

The analytical calculations yield 
the following expression for the transmission:\cite{Muj94b} 
\begin{widetext}
\begin{equation}
\label{Tg}
T(E)= 
4\xi^{2}\left|\frac{\sin\{\theta\}}{\sin\{(N+1)\theta\}+
2i\xi\sin\{ N\theta\}-\xi^{2}\sin\{(N-1)\theta\}}\right|^{2}.
\end{equation}
\end{widetext}
where we have introduced the dimensionless quantities 
\begin{equation}
\cos\theta\equiv\frac{E-E_{0}}{2V},\,\,\,\xi\equiv\frac{\Delta}{V}.
\label{par}
\end{equation}

 If $|\frac{E-E_{0}}{2V}|>1$, then the trigonometric function in
eqs.(\ref{Tg},\ref{par}) are transformed into the corresponding
hyperbolic functions. This results in the exponential scaling of the transmission with 
the nanowire length $N$.\cite{Muj94b} If $|\frac{E-E_{0}}{2V}|\leq1$,
then eq.(\ref{Tg}) describes the resonance tunneling regime where
the transmission oscillates as a function of the nanowire length $N$.
Oscillatory regime is the intrinsic property of the single channel
atomic nanowires such as Au and Na,\cite{sim01,tsuk02,lee04,khom04,lang98,g_osc}
to which the one-dimensional tight-binding Hamiltonian (\ref{H}) is applicable.
Therefore we restrict our consideration to this regime and demonstrate
that the transmission exhibits scaling and self-similarity.

Suppose that the coupling between the wire and the lead is weak, i.e.
$\xi<1$. The transmission $T(E)\sim\xi^{2}+O(\xi^{3})$ is very small
for $\xi\ll1$ unless $\sin\{(N+1)\theta\}$ in the denominator of
eq.(\ref{Tg}) tends to zero. If we put $\sin\{(N+1)\theta\}=0$,
which is equivalent to assuming that 
\begin{equation}
\theta_{k}=\frac{\pi k}{N+1},\,\, E_{k}=E_{0}+2V\cos\left\{ \theta_{k}\right\} ,\,\, k=1,2,...,N
\label{res}
\end{equation}
 then the transmission achieves its maximum \begin{equation}
T(E_{k})=\frac{1}{1+\xi^{2}\cos^{2}\left\{ \theta_{k}\right\} }.\label{Tmax}\end{equation}
 For any given length of the wire there are $N$ different resonances
(\ref{Tmax}) in the transmission. The resonance label $k$ in eqs.(\ref{res}),
 (\ref{Tmax}) can be interpreted as the number of the transmission maxima
achieved for the given molecular length $N$ within the energy interval
$[E_{k},E_{0}+2V]$.

The resonance condition (\ref{res}) suggests the existence of the
self-similarity of the transmission $T(E)$ as we vary the nanowire
length $N$ and energy $E$ of transmitted electrons. Indeed, the following simple linear transformation
\begin{equation}
\tilde{N}=j(N+1)-1,\label{mapN}\end{equation}
\begin{equation}
\tilde{\theta}=\theta/j,\,\, j>1\label{mapTet}\end{equation}
 leaves the resonance condition (\ref{res}) unaffected, thereby preserving
the number of resonances of the transmission by deforming the energy
interval. The mapping parameter $j=(\tilde{N}+1)/(N+1)$ is not, in
general, an integral but a rational number. If the angles $\theta$
and $\tilde{\theta}$ are expressed explicitly in terms of the corresponding
energies $E$ and $\tilde{E}$ via eq.(\ref{par}), then the transformation
(\ref{mapTet}) becomes nonlinear: \begin{equation}
\tilde{E}=E_{0}+2V\cos(\arccos\left\{ \frac{E-E_{0}}{2V}\right\} /j),\,\,\, j>1.\label{mapE}\end{equation}
The transformation (\ref{mapN}), (\ref{mapE}) leads to the self-similarity
of transmission at $[\tilde{E},E_{0}+2V]$ and $[E,E_{0}+2V]$ energy
intervals: the wire of the length $\tilde{N}$ within the interval
$[\tilde{E},E_{0}+2V]$ has exactly the same number $k$ of
the transmission maxima as the wire of the length $N$ within the
interval $[E,E_{0}+2V]$.
Furthermore, consider the transmission $T(E)$ of a wire of
the length $N$ when the energy of incident electrons
varies within the range  $[E_{1},E_{2}]$. Then $T(E)$
can be mapped into $\tilde{T}(E)$ for any $\tilde{N}>N$ within the
window $[\tilde{E_{1}},\tilde{E_{2}}]$, provided that $j$ is defined
via eq. (\ref{mapN}). The number and positions of the conductance
peaks (\ref{res}) for $T(E)$ within $[E_{1},E_{2}]$ will be exactly
the same as that of $\tilde{T}(E)$ within $[\tilde{E_{1}},\tilde{E_{2}}]$.

\section{Example calculations}

\subsection{Numerical calculations with one-dimensional tight-binding model}

We confirm the existence of the self-similarity of a single-channel transmission
by numerical calculations via eq.(\ref{Tg}).
The computations are performed for $E_{0}=0$ eV, $V=2.5$
eV and $\xi=0.04$. Several illustrative results are presented in
Fig.1 for rational mapping parameters $j$. The transmission $T(E)$
is computed within the energy windows calculated as prescribed
by eqs.(\ref{mapN}) and (\ref{mapE}). The transmissions for $N=21,\,34$
and $107$ exhibit almost perfect self-similarity. Note that the scale
of the vertical axes in the figures is not the same. This means
that the transmission curves should be additionally stretched in order
to make a perfect match between them.
 Fig. 1 shows
that the widths of the peaks are broader for shorter nanowires. This
has the following explanation. If we expand transmission (\ref{Tg}) in the
vicinity of $E\approx E_{k}$ and retain the leading on $E-E_{k}$
and $\xi$ contributions, we obtain: 
\begin{equation}
T(E)\approx\frac{1}{1+\left[\frac{NV(E-E_{k})}{\xi\{4V^{2}-(E_{k}-E_{0})^{2}\}}\right]^{2}}\label{shape}
\end{equation}
 Thus, the higher is $N$ and/or $k$, the narrower is the Lorentzian
peak-shape (\ref{shape}).

\begin{figure}
\centerline{
\epsfig{figure=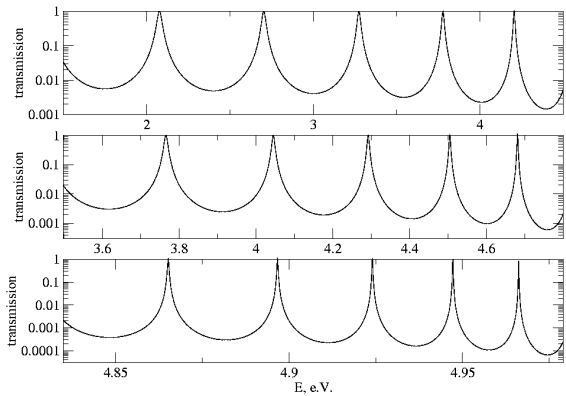,width=\columnwidth,angle=-0}}
\caption{Transmission vs. energy for $N=21$ (upper panel), $34$ (middle
panel), and $107$ (bottom panel). For $N=21$, the energy window
$1.5\leq E\leq4.5$  has been chosen. For $N=34$ ($j=1.59$)
and $107$ ($j=4.91$) the corresponding energy windows are $3.50\leq E\leq4.80$
and $4.83\leq E\leq4.98$  } 
\label{fig:fig1}
\end{figure}

The procedure described above opens the possibility of two-dimensional
continuation of transmission on the $E-N$ plane. The transmission
$T(E)$ in the rectangular domain $[E_{1}<E<E_{2}]$, $[N_{1}<N<N_{2}]$
can be mapped into that of the domain $[\tilde{E_{1}}<E<\tilde{E_{2}}]$,
$[\tilde{N}_{1}<N<\tilde{N}_{2}]$ via the transformation (\ref{mapN},
\ref{mapE}).

Most tight-binding and first principle transport calculations relay on the nonorthogonal 
atomic basis functions to compute the wire Green's functions.\cite{xue02} 
Within the non-orthogonal representation the Landauer formula (\ref{T}) 
remains valid, but the definition
of the Green's function (\ref{Gr}) should be  modified according to 
\begin{equation}
\mathbf{G}(E)=[E\mathbf{S}-\mathbf{H}-\mathbf{\Sigma}(E)]^{-1},
\label{GrS}
\end{equation}
where $\mathbf{S}$ is the overlap matrix. If we assume that
only the nearest-neighbor basis functions overlap significantly, then
the matrix $\mathbf{S}$ reduces to the three-diagonal form:
\begin{equation}
\mathbf{S}=\left[\begin{array}{ccccccc}
1 & \lambda & 0 & 0 & 0 & 0 & 0\\
\lambda & 1 & \lambda & 0 & 0 & 0 & 0\\
0 & \lambda & 1 & \lambda & 0 & 0 & 0\\
0 & 0 & . & . & . & 0 & 0\\
0 & 0 & 0 & . & . & . & 0\\
0 & 0 & 0 & 0 & \lambda & 1 & \lambda\\
0 & 0 & 0 & 0 & 0 & \lambda & 1\end{array}\right],
\label{S}
\end{equation}
where $|\lambda|<1$ is the overlap integral between the nearest neighbor basis functions.
In this case, the expression for the transmission (\ref{Tg}) remains correct if 
the parameter $\theta$ is redefined: 
\begin{equation}
\cos\theta\equiv\frac{E-E_{0}}{2(V-E\lambda)}.
\label{parS}
\end{equation}
Therefore, the non-orthogonality of the basis functions breaks the
mirror symmetry of $T(E)$ with respect to the sign of the site energy $E_{0}$,
and different choices of the site energy $E_{0}$ are no longer equivalent. 

Having adopted new definition (\ref{parS}) for the parameter $\theta$,
we observe that the self-similarity transformation (\ref{mapN},
\ref{mapTet}) remains unchanged. The counterpart of the transformation
(\ref{mapE}) in the energy domain should be modified to account for the 
basis nonorthogonality: 
\begin{equation}
\tilde{E}=\frac{E_{0}+2V\cos(\arccos\left\{ \frac{E-E_{0}}{2(V-E\lambda)}\right\} /j)}
{1+2\lambda\cos(\arccos\left\{ \frac{E-E_{0}}{2(V-E\lambda)}\right\} /j)},\,\,\, j>1.
\label{mapES}
\end{equation}
 If $\lambda=0$, then the transformations (\ref{mapE}) and (\ref{mapES})
coincide, evidently. If $\lambda\neq0$, the non-orthogonality of
the basis functions causes, through the (energy-dependent) scaling
factor $(1+2\lambda\cos(\theta/j))^{-1}$, an additional stretching
(shrinking) in the energy space, depending on whether the product
$\lambda\cos(\theta/j)$ is negative (positive). For example, if we
assume that the overlap parameter $\lambda=0.2$ and plot the transmissions
analogies to those in Fig. 1, we obtain exactly the same curves. However,
according to eqs.(\ref{mapN}) and (\ref{mapES}), the energy intervals
will shrink (we use eV units for energy $E$): $1.34\leq E\leq3.31$ for $N=21$, $2.73\leq E\leq3.47$
for $N=34$, and $3.49\leq E\leq3.56$  for $N=107$. On
the contrary, if we put $\lambda=-0.2$ (this is equivalent to the
consideration of $T(-E)$ for $\lambda=0.2$) and plot the same transmissions,
the corresponding energy intervals will stretch as compared with those
in Fig. 1: $1.70\leq E\leq7.03$  for $N=21$, $4.86\leq E\leq7.79$
for $N=34$, and $7.88\leq E\leq8.27$  for $N=107$.

\subsection{First principle simulations for sodium nanowires}

In this section, we examine the prediction of transmission self-similarity
by performing the first principle simulations of transport properties of Na atomic wires.
The calculations were performed using our plane wave/pseudopotential implementation of 
the density functional based non-equilibrium Green's function techniques in the 
 CPMD package.\cite{li05,cpmd} 
Our implementation is based on  nonorthogonal
Wannier-type atomic orbitals.\cite{li05} 
All  systems 
were treated employing periodic boundary conditions and the Kohn-Sham orbitals 
were expanded in plane waves (50 Ry cutoff) at the $\Gamma$ point of the Brillouin zone. 
We used local density approximation for the exchange-correlation functional 
and Stumpf, Gonze, and Schettler 
pseudopotentials\cite{gonze9103} for core electrons. 
The system is treated as a ``supermolecule'' placed in a large supercell. The size of  supercell  
is chosen in such a way that the distance between the nearest atoms in the 
neighboring cells is larger  
than 8.5\AA, so that the interaction between supercell images is negligible.  
The whole system is divided into three parts: left electrode, central wire, and right electrode. 
 The electrode part is obtained by cutting a few atoms from Na (001) surface.  The geometry 
of the electrodes is fixed to the bulk values. The wire part is a  
single chain of Na atoms, where the distance between the atoms is constrained to  
the nearest neighbor distance in the bulk system. The distance between the electrode  
part and the wire part is optimized.
 
The results of our first principle transport calculations are illustrated by Figs. 2.
Figs. 2a and 2b depict the transmission
calculated for Na nanowire with $N=3$ and $5$, respectively. Figs. 2c and 2d show
zoomed portions of Figs. 2a and 2b, which are seen to exhibit quite
a good resemblance.

To link these calculations to the theory developed in the paper,
the parameters of the tight-binding molecular Hamiltonian
 $\mathbf{H}$,  the self-energy matrix $\mathbf{\Sigma}(E)$ and the overlap matrix $\mathbf{S}$
were extracted from the first principle transport calculations. 
Since we use nonorthogonal basis set,
the Green's function should be taken in the form (\ref{GrS}) which accounts for the overlap matrix
$\mathbf{S}$ between Wannier type atomic orbitals.
The calculated molecular Hamiltonian and overlap
matrix are described fairly well by the three-diagonal matrices (\ref{H}) and (\ref{S}),
respectively, with the parameters $E_{0}=-2.88$ eV, $V=1.25$ eV, 
and $\lambda=-0.28$. If we insert these parameters into eqs.
(\ref{mapN}) and (\ref{mapES}), then the energy window $[-0.56,0]$
for $N=3$ will be mapped into $[-0.57,-0.35]$ for $N=5$. This
correlates reasonably well with the actual window $[-0.64,-0.35]$.
Of course, the calculated $\mathbf{H}$ and $\mathbf{S}$ matrices
comply with simple three-diagonal formulas (\ref{H}) and (\ref{S}) only approximately,
and the actual self-energy function $\Sigma(E)=\Lambda(E)-i\Delta(E)$
is neither purely imaginary, nor energy independent. Evidently, the
presence of $\Lambda(E)\neq0$ causes an additional shift of transmission
maxima, and $\Sigma(E)$ brings an additional energy dependence of
$T(E)$ which has not been taken into account within the present theoretical
analyses.  If, however, $\Sigma(E)$ does not change significantly within 
the interval$[E_{1}<E<E_{2}]$, then the self-similarity in $T(E)$ should be
preserved, as is proven by Figs. 2c and 2d. 

\begin{figure}
\centerline{
\epsfig{figure=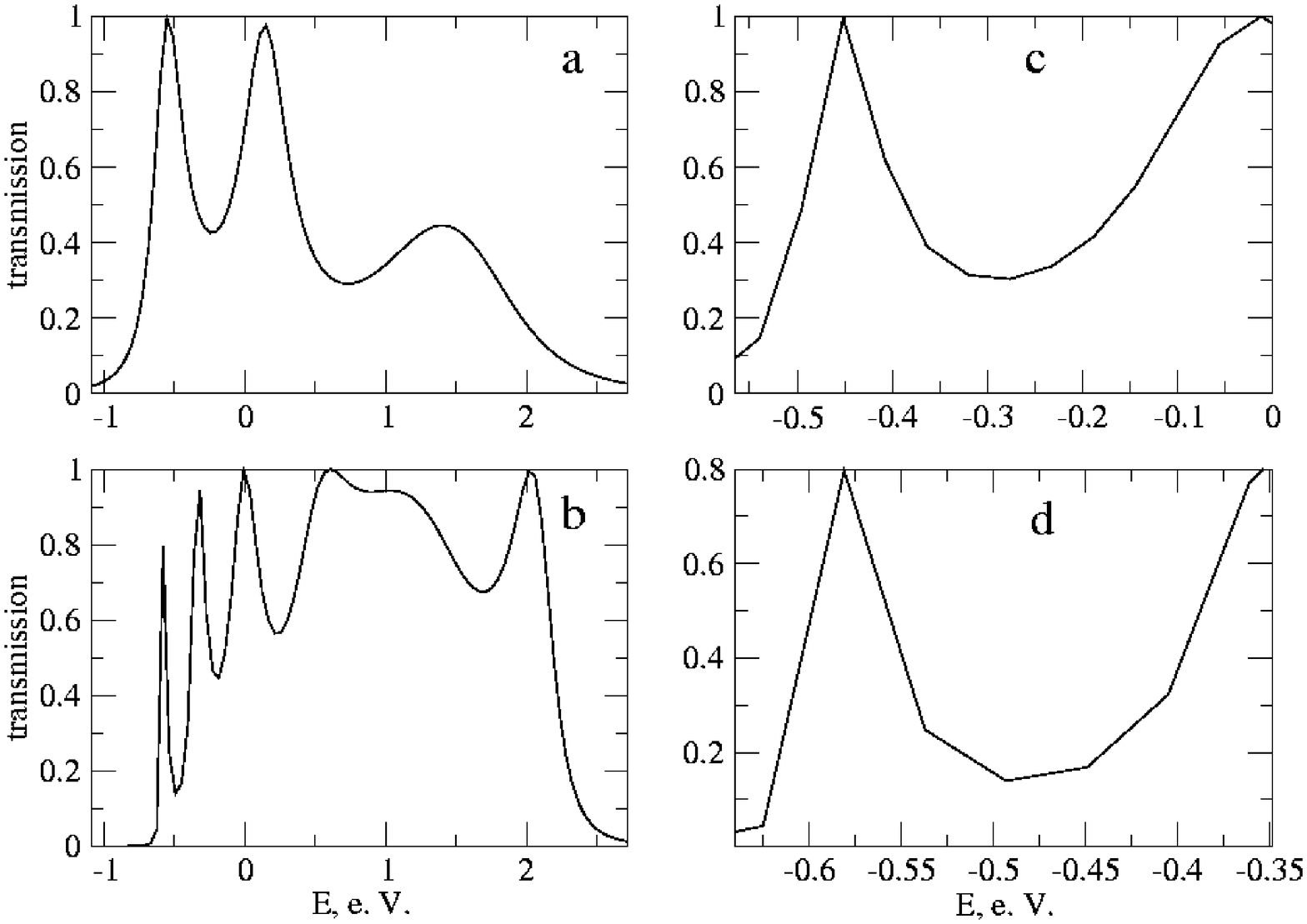,width=\columnwidth,angle=-0}}
\caption{
First principle simulations for Na nanowire. 
Transmissions calculated for $N=3$ and $N=5$ are presented in Figs. (a) and (b), respectively.
Fig. (c) depicts a portion of Fig. (a) at the energy interval
$[-0.56,0]$, and Fig. (d) shows a part of Fig. (b) at the interval
$[-0.64,-0.35]$. The Fermi energy is set to zero.
} 
\label{fig:fig2}
\end{figure}

\section{Discussion and Outlook}

In this paper, we have demonstrated that a single-channel transmission  
for electron transport in nanowires exhibits the self-similarity in resonance tunneling regime.
Starting from the resonance condition, we proposed the transformation of  
the wire length and the incident electron energy,
which preserves the number of maxima in the transmission. 
A sample calculations on one-dimensional tight binding model produces 
exact agreement with analytical results. The density functional based transport calculations
of Na atomic wires confirm the existence of self-similarity in the transmission for realistic system.

The existence of scaling properties and self-similarity of the molecular
transmission is also supported by a possibility to define it via
the Friedel sum rule \cite{Datt_97}. It is not necessary to have a detailed
information on a large variety of the parameters describing the molecule
and the leads to predict transport properties of a single channel conductor. 
To compute the single-channel conductance by the Friedel sum rule, it is sufficient to know
only the difference between the number of electrons occupying the even
and the odd eigenstates of the molecule. 

It is thus not very unrealistic to  suggest that the self-similarity
in molecular transmission and conductance may be a rather general
phenomenon. In terms of possible applications, our results suggest
that the transmission, computed within a certain energy and length
window, can be transferred to other windows via the transformation
(\ref{mapN}), (\ref{mapE}), (\ref{mapES}). For example, the transmission
of a short wire within a relatively wide energy window can be mapped
into the transmission of a longer wire within a narrower energy window.
If we assume, additionally, that the site energy $E_{0}$ can be purposefully
modified,\cite{All_03,Datta_04} then the molecular conductance $g$,
eq.(\ref{g}), becomes a function of the (modified) site energy $E_{0}$
and molecular length $N$ and also exhibits the self-similarity.

\begin{acknowledgments}
We are grateful to Ferdinand Evers and Michael Thoss for fruitful
and stimulating discussions. 
\end{acknowledgments}


\begin{thebibliography}{40}

\bibitem{agrait03}N. Agrait, A.L. Yeyati, J. M. van Ruitenbeek, Phys. Rep. 377, 81 (2003). 

\bibitem{datta04}S. Datta, Nanotechnology 15, S433 (2004). 

\bibitem{nitzan03}A. Nitzan, M.Ratner, Science 300, 1384 (2003).


\bibitem{Mer_85}R. Merlin, K. Bajema, and R. Clarke, Phys. Rev. Lett. 55, 1768 (1985).

\bibitem{Cott_03}E. L. Albuquerque and M.G. Cottam, Phys. Rep. 376, 225 (2003).

\bibitem{Gap_02}A. V. Lavrinenko, S. V. Zhukovsky, K. S. Sandomirski, and S. V. Gaponenko,
Phys. Rev. B 65, 036621 (2002). 

\bibitem{Gap_04}S. V. Zhukovsky, A. V. Lavrinenko, and S. V. Gaponenko, Europhys.
Lett. 66, 455 (2004). 

\bibitem{Alb_99}M.S. Vasconcelos and E. L. Albuquerque, Phys. Rev. B 59, 11128 (1999).

\bibitem{Gong_98}X. Huang and C. Gong, Phys. Rev. B 58, 739 (1998).

\bibitem{Mac_97}E. Macia, Phys. Rev. B 57, 7661 (1997).

\bibitem{Roy_95a}C. L. Roy and A. Khan, Phys. Lett. A 196, 346 (1995).

\bibitem{Roy_95b}C. L. Roy, A. Khan, and C. Basu, J. Phys.: Condens. Matter 7, 1843
(1995).

\bibitem{Sal_98}W. Salejda and P. Szyszuk, Physica A 252, 547 (1998).

\bibitem{Sal_02}M. H. Tyc and W. Salejda, Physica A 303, 493 (2002).

\bibitem{Wang_00}R. Ovideo-Roa, L. A. Perez, and C. Wang, Phys. Rev. B 62, 13805 (2000).

\bibitem{Wang_01}V. Sanchez, L. A. Perez, R. Ovideo-Roa, and C. Wang, Phys. Rev. B
64, 174205 (2001).

\bibitem{Claro_02}Z. Y. Zeng and F. Claro, Phys. Rev. B 65, 064207 (2002).

\bibitem{2D_bil}A. P. Micolich, R. P. Taylor, T. P. Martin, R. Newbury, T. M. Fromhold,
A. G. Davies, H. Linke, W. R. Tribe, L. D. Macks, C. G. Smith, E.
H. Linfield, and D. A. Ritchie, Phys. Rev. B 70, 085302 (2004).
 
\bibitem{per02}P. Pereyra and E. Castillo, Phys. Rev. A 65, 205120 (2002).

\bibitem{Oni_99}A. Onipko, Y. Klymenko, and L. Malysheva, Material Science and Engineering
C 8-9,  273 (1999).

\bibitem{Muj94a}V. Mujica, M. Kemp, M.A. Ratner, J. Chem. Phys. 101, 6849 (1994). 

\bibitem{Muj94b}V. Mujica, M. Kemp, M.A. Ratner, J. Chem. Phys. 101, 6856 (1994). 

\bibitem{sim01}H. -S. Sim, H. -W. Lee, and K. J. Chang, Phys. Rev. Lett. 87, 96803
(2001).

\bibitem{tsuk02}S. Tsukamoto and K. Hirose, Phys. Rev. B 66, 161402 (2002).

\bibitem{lee04}Y. J. Lee, M. Brandbyge, M. J. Puska, J. Taylor, K. Stokbro, and R.
M. Nieminen, Phys. Rev. B 69, 125409 (2004).

\bibitem{khom04}P. A. Khomyakov and G. Brocks, Phys. Rev. B 70, 195402 (2004).

\bibitem{lang98}N. D. Lang and Ph. Avouris, Phys. Rev. Lett. 81, 3515 (1998). 

\bibitem{g_osc}R. H. M. Smit, C. Untiedt, G. Rubio-Bollinger, R. C. Segers, and J.
M. van Ruitenbeek, Phys. Rev. Lett. 91, 076805 (2003).

\bibitem{xue02}Y. Xue, S. Datta, M. A. Ratner, Chem. Phys. 281, 151 (2002).

\bibitem{li05} Z. Li and D.S. Kosov, cond-mat/0507649

\bibitem{cpmd} 
CPMD, Copyright IBM Corp 1990-2001, Copyright MPI f\"ur Festk\"orperforschung Stuttgart 1997-2004.

\bibitem{gonze9103} X. Gonze, R.Stumpf, and M.Scheffler, Phys. Rev. B 44, 8503 (1991).

\bibitem{Datt_97}S. Datta and W. Tian, Phys. Rev. B 55, R1914 (1997). 

\bibitem{All_03}V. Mujica, A. Nitzan, S. Datta, M. A. Ratner, C. P. Kubiak, J. Phys.
Chem. B 107, 91 (2003).

\bibitem{Datta_04}A. W. Ghosh, T. Rakshit and S. Datta, Nano Lett. 4, 565 (2004).

\end{thebibliography}
\end{document}